\documentclass{XrU2005}

\usepackage{psfig,color,pstricks}
\title{Ionization structure of the warm wind in NGC~5548}
\author[1,2]{K. C. Steenbrugge}
\author[2]{J. S. Kaastra}
\author[3]{D. M. Crenshaw}
\author[4,5]{S. B. Kraemer}
\author[6]{N. Arav}
\author[7,8]{I. M. George}
\author[9]{D. A. Liedahl}
\author[10]{F. B. S. Paerels}
\author[7,8]{T. J. Turner}
\author[8,11]{T. Yaqoob}
\affil[1]{SRON National Institute for Space Research, Sorbonnelaan 2,
3584 CA Utrecht, The Netherlands}
\affil[2]{now at Harvard-Smithsonian Center for Astrophysics, 60 Garden street,
Cambridge, MA 02138, USA}
\affil[3]{Department of Physics and Astronomy, Georgia State University, Astronomy Offices, One Park Place South SE, Suite 700, Atlanta, GA 30303, USA}
\affil[4]{Catholic University of America, USA}  
\affil[5]{Laboratory for Astronomy and Solar Physics, NASA's Goddard
Space Flight Center, Greenbelt, MD 20771, USA}
\affil[6]{CASA, University of Colorado, 389 UCB, Boulder, CO 80309-0389, USA}
\affil[7]{Joint Center for Astrophysics, University of Maryland
Baltimore County, 1000 Hilltop Circle, Baltimore, MD 21250, USA}  
\affil[8]{Exploration of the Universe Division, Code 662, NASA's
Goddard Space Flight Center, Greenbelt, MD 20771, USA}
\affil[9]{Physics Department, Lawrence Livermore National Laboratory, PO Box 808, L-41, Livermore, CA 94550, USA}
\affil[10]{Columbia Astrophysics Laboratory, Columbia University, 538W. 120th Street, New York, NY 10027, USA}
\affil[11]{Department of
Physics and Astronomy, Johns Hopkins University, Baltimore, MD 21218,
USA}

\begin{document}


\maketitle

\begin{abstract}
We present the results from our 140~ks XMM-{\it Newton} and 500~ks {\it
Chandra} observation of NGC~5548. The velocity structure of the X-ray absorber is consistent with the
velocity structure measured in the simultaneous UV
spectra. In the X-rays we can
separate the highest outflow velocity component, $-$1040 km s$^{-1}$,
from the other velocity components. This
velocity component spans at least three orders of magnitude in
ionization parameter, producing both highly ionized X-ray absorption
lines (Mg XII, Si XIV) and UV absorption lines. A similar
conclusion is very probable for the other four velocity components.  
We show that the lower ionized absorbers are not in pressure
equilibrium with the rest of the absorbers.
Instead, a model with a continuous distribution of column density
versus ionization parameter gives an excellent fit to our data. 
\end{abstract}

\section{Introduction}

 Over half of all Seyfert 1 galaxies exhibit signatures of
photoionized outflowing gas in their X-ray and UV spectra. Studying
these outflows is important for a better understanding of the
enrichment of the Inter Galactic Medium (IGM) as well as the physics
of accretion of gas onto a super-massive black hole. 
Arav et al. \citep{arav03} and Steenbrugge et al. \citep{steenbrugge03} concluded that there is substantially more lowly ionized gas than has been claimed from previous UV observations. It was concluded that the X-ray and UV warm absorbers are different manifestations of the same phenomenon.

\section{Observation and data reduction}

The XMM-{\it Newton} RGS data was reduced using the standard threads
of the SAS version 5.3. For the {\it Chandra} HETGS data we used the
threats in CIAO version 2.2. For the LETGS the 1.5 event file was
obtained using CIAO version 2.2, but further data reduction was done
using the pipeline described by \citet{kaastra02}, which includes an empirical correction for the 
known wavelength problem in the LETGS and fitted it with responses that include the 
first 10 positive and negative orders. The data were analyzed using
the {\it spex} package \citep{kaastra02b}.

\section{Velocity structure}

\begin{table}[!h]
\begin{center}
\caption{The outflow velocity, the velocity broadening and ionization
parameter as measured for the five components detected in the UV
\citep{crenshaw03}.}
\begin{tabular}{l|c|l|l}\\\hline
Outflow & broadening & U    & N$_{\rm H}$ \\
km s$^{-1}$ & km s$^{-1}$ & & log m$^{-2}$ \\\hline
$-$1040   &  94  & 0.03     & 22.8 \\
$-$667    &  18  & 0.03     & 22.6 \\
$-$530    &  68  & 0.24     & 24.3 \\
$-$336    &  62  & 0.03     & 23.3 \\
$-$160    &  90  & 0.03     & 22.6 \\\hline
\end{tabular}
\end{center}
\end{table}

\begin{figure}
\centering
\psfig{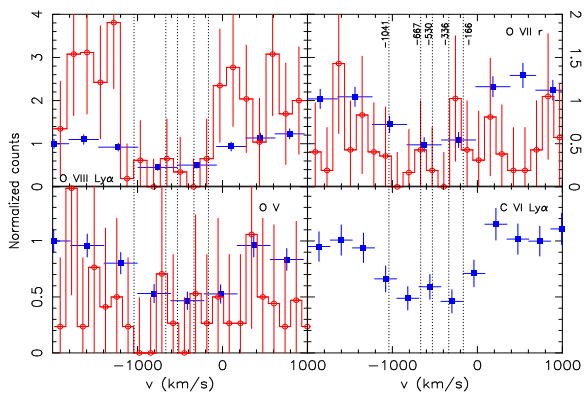}
\caption{The MEG (open circles) and LETGS (filled squares) data for
the O~VIII~Ly$\alpha$, O~VII resonance, O~V and C~VI Ly$\alpha$ lines. The dotted lines indicate the outflow velocity measured from UV 
spectra.}
\label{fig:vel}
\end{figure}

The five velocity components measured in the UV \citep{crenshaw03} are 
listed in Table 1. Each velocity component can have a
different ionization parameter and hydrogen column density. Also the
variability detected in the ionization parameter is different for the
five outflow velocities.  
Using the MEG, HEG and LETGS data we
were able to resolve the -1040 km s$^{-1}$ component from the 4 other
velocity components in the 6 strongest lines (see Fig~\ref{fig:vel}). This clearly indicates
that this velocity component spans an ionization range of at least 3
orders of magnitude, from low ionization UV lines to Si~XIV.

\section{Ionization structure}
\begin{figure}
\centering
\psfig{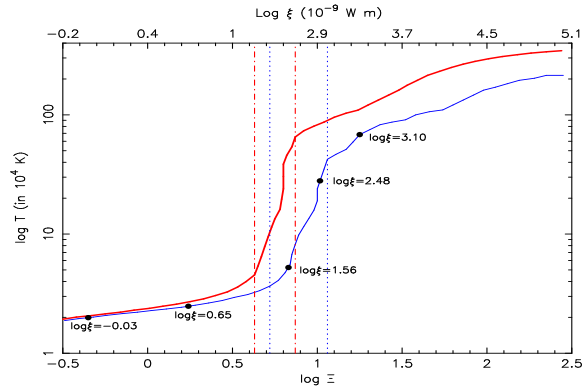}
\caption{The temperature versus ionization parameter for constant
pressure. Points: the ionization paramters measured from the RGS 
spectra. The ionization parameter $\xi$ for the spectral energy
distribution (SED) assumed in the upper curve is indicated on the top
x-axis.  The dotted and dashed lines
indicate the boundaries for the marginally stable branch for the two
different SEDs assumed.}
\label{fig:ion}
\end{figure}

Fig.~\ref{fig:ion} shows the S-curve, with superimposed the five ionization
parameters measured from the RGS spectra. The two lowest ionized components measured 
from the X-ray spectra cannot be in pressure equilibrium with the
higher ionized components. This means that if the absorber is due to
clouds in an outflow, they need to be magnetically confined. An equally good fit to the spectra is obtained with a continuous
ionization parameter distribution, i.e. an outflow with a density
gradient \citep{steenbrugge05}. The lowest ionization component measured in 
the X-rays has a very similar ionization parameter as the ionization
parameter measured from UV spectra: log $\xi_{\rm X-ray}$ = $-$ 0.03
versus log $\xi_{\rm UV}$ = 0.05. However, most of the gas is highly ionized.
Assuming a continuous outflowing stream, we derive a power law slope
for the column density of 0.4 with ionization parameter.

\section{Acknowledgements}
SRON National Institute for Space Research is supported financially by 
NWO, the Netherlands Organization for Scientific Research.

\bibliography{/data/guido/katrien/cfaarticle/references}

\begin{thebibliography}{6}
\expandafter\ifx\csname natexlab\endcsname\relax\def\natexlab#1{#1}\fi

\bibitem[{{Arav} {et~al.}(2003){Arav}, {Kaastra}, {Steenbrugge}, {Brinkman},
  {Edelson}, {Korista}, \& {de Kool}}]{arav03}
{Arav}, N., {Kaastra}, J., {Steenbrugge}, K., {et~al.} 2003, \apj, 590, 174

\bibitem[{{Crenshaw} {et~al.}(2003){Crenshaw}, {Kraemer}, {Gabel}, {Kaastra},
  {Steenbrugge}, {Brinkman}, {Dunn}, {George}, {Liedahl}, {Paerels}, {Turner},
  \& {Yaqoob}}]{crenshaw03}
{Crenshaw}, D.~M., {Kraemer}, S.~B., {Gabel}, J.~R., {et~al.} 2003, \apj, 594,
  116

\bibitem[{{Kaastra} {et~al.}(in press){Kaastra}, {Mewe}, \&
  {Raassen}}]{kaastra02b}
{Kaastra}, J.~S., {Mewe}, R., \& {Raassen}, A.~J.~J. in press, Proc. Symp. New
  Visions of the X-ray Universe in the XMM-{\it Newton} and {\it Chandra} era

\bibitem[{{Kaastra} {et~al.}(2002a){Kaastra}, {Steenbrugge}, {Raassen}, {van
  der Meer}, {Brinkman}, {Liedahl}, {Behar}, \& {de Rosa}}]{kaastra02}
{Kaastra}, J.~S., {Steenbrugge}, K.~C., {Raassen}, A.~J.~J., {et~al.} 2002a,
  \aap, 386, 427

\bibitem[{{Steenbrugge} {et~al.}(2005){Steenbrugge}, {Kaastra}, {Crenshaw},
  {Kraemer}, {Arav}, {George}, {Liedahl}, {van der Meer}, {Paerels}, {Turner},
  \& {Yaqoob}}]{steenbrugge05}
{Steenbrugge}, K.~C., {Kaastra}, J.~S., {Crenshaw}, D.~M., {et~al.} 2005, \aap,
  434, 569

\bibitem[{{Steenbrugge} {et~al.}(2003){Steenbrugge}, {Kaastra}, {de Vries}, \&
  {Edelson}}]{steenbrugge03}
{Steenbrugge}, K.~C., {Kaastra}, J.~S., {de Vries}, C.~P., \& {Edelson}, R.
  2003, \aap, 402, 477

\end{thebibliography}

\end{document}